\documentstyle[preprint,aps]{revtex}
\renewcommand{\theequation}{\arabic{section}.\arabic{equation}}
\tighten
\begin{document}
\draft 
\title{A Non-perturbative Treatment of Heavy Quarks and Mesons} 
\author{C.\ J.\ Burden\footnotemark[1] and 
D.-S. Liu\footnotemark[2]\vspace*{0.2\baselineskip}}
\address{
\footnotemark[1]Department of Theoretical Physics,
Research School of Physical Sciences and Engineering,
Australian National University, Canberra, ACT 0200, 
Australia\vspace*{0.2\baselineskip}\\
\footnotemark[2]Department of Physics,
University of Tasmania,
GPO Box 252C, Hobart, Tasmania 7001, Australia\vspace*{0.2\baselineskip} }
\maketitle 
\begin{abstract}
A formalism for studying heavy quarks in terms of model 
Dyson-Schwinger equations is developed.  The formalism is the natural 
extension of a technique which has proved successful in a number of 
studies of light hadron physics.  The dressed heavy quark propagator, 
calculated to leading order in the inverse quark mass, is incorporated 
in a treatment of mesons consisting of a heavy quark and light antiquark 
via the ladder approximation Bethe-Salpeter equation.  In the limit of 
infinite heavy quark mass the model is found to respect the spectrum 
degeneracies present in Heavy Quark Effective Theory. An exploratory 
numerical analysis of a simple form of the model is carried out to 
assess its viability for studying $D$ and $B$ mesons.
\end{abstract} 
\section{Introduction}

There has been a great deal of recent theoretical interest in heavy 
quark mesons, that is, mesons containing charm quarks or 
bottom quarks such as $B$ and $D$ mesons.  The 
basis of this interest is heavy quark effective theory (HQET) \cite{N94}, 
which hinges on the realisation that in the limit of infinite quark mass 
extra symmetries, and hence degeneracies, occur beyond those normally 
associated with QCD \cite{S80}.  It is a property of many existing HQET 
analyses, particularly of meson spectrum calculations, that dynamical 
self dressing of the quarks is either not included at all, or rarely 
included in a non-perturbative fashion.  Examples of recent spectrum 
calculations in which heavy quarks are modelled essentially by free 
propagators are given in ref. \cite{Z95}.  However, one could argue that 
a theory  which does not prevent quarks from being on mass shell is 
not in keeping with the spirit of a confinement and may lead, in 
certain instances, to incorrect results.  In Bethe-Salpeter equation 
calculations for instance, there is in principle nothing to prevent the 
occurrence of spurious production thresholds.  

At the other end of the meson spectrum, namely the physics of light quarks, 
the Dyson-Schwinger equation (DSE) 
technique \cite{RW94} has enjoyed considerable success in modelling 
a number of light meson phenomena.  
Basic to the DSE technique is the quark propagator.  The spacelike 
structure of this propagator has been well modelled within the DSE 
technique for light quarks.  At a phenomenological level, model quark 
propagators based on numerical studies of approximate Dyson Schwinger 
equations have proved efficacious in calculating light meson 
spectra \cite{BSp1,BSpapers,C95}, 
electromagnetic form factors of the pion and kaon \cite{R95}, $\pi$-$\pi$ 
scattering lengths \cite{R94}, quark loop contributions to $\rho$-$\omega$ 
mixing \cite{M94} and the anomalous $\gamma\pi \rightarrow \gamma$ and 
$\gamma \pi \rightarrow \pi\pi$ form factors \cite{F95}.  

The aim of this paper is to formulate an extension of the DSE technique 
to the realm of heavy quarks.  In the applications of the the DSE 
technique to light quarks listed above, 
it has not been necessary to determine the behaviour of quark propagators 
well away from the real spacelike $p^2$-axis.  However, for the purpose of 
accurately modelling confined heavy quarks, because the self energy is 
small compared with the current quark mass $m_Q$, it is particularly important 
to know the analytic structure of the heavy quark propagator near the 
(timelike) bare fermion mass pole $p_{\rm Minkowski}^2 = m_Q^2$.  
The method frequently used within the DSE technique of fitting 
analytic functions to the 
numerically determined spacelike quark propagator is inadequate for this 
task.   It is for this reason that new ideas must be developed within the DSE 
technique if it is to be adapted to heavy quarks.  

It is the philosophy of the DSE  
technique that confinement in QCD should be signalled by the absence of 
timelike poles in the quark propagator.  This analytic structure is the 
result of non-perturbative quark self dressing, which in principle 
could be demonstrated by a full treatment of the quark Dyson-Schwinger 
equation.  Many numerical calculations of approximate quark 
Dyson-Schwinger equations in confining field theories have demonstrated 
propagator solutions which are free of timelike poles \cite{RW94}, 
though not necessarily free of complex conjugate poles away from the 
real $p^2$-axis \cite{Ma95}.  Herein we shall examine in detail 
the analytic structure of the heavy quark propagator in the limit of 
infinite quark mass.  We shall also consider the Bethe-Salpeter equation 
(BSE) for bound states of heavy quarks and light antiquarks.  Our eventual 
aim is to make contact with the observed spectrum of $D$ and $B$ mesons.  
However, at the present level of approximation we are unable to find 
solutions to our model BSE because of spurious poles in the model quark 
propagators.  We suggest ways in which the model can be improved to 
overcome this problem.  

The layout of the paper is as follows: 
In Section II we develop the basic formalism for the heavy quark propagator.  
In Section III we consider the Bethe-Salpeter equation for 
mesons consisting of one heavy and one light antiquark.  Sections IV and V 
deal with scalar and vector mesons respectively, and demonstrate that the 
symmetries of HQET are preserved within the heavy quark limit of the DSE 
technique.  Some numerical results are given in Section VI and conclusions 
drawn in Section VII.  

\setcounter{equation}{0}
\section{The heavy quark propagator.}

Before understanding the procedures which must be 
employed to ensure the extension of the DSE technique to heavy quarks, 
it is important 
to appreciate that a naive translation of the light quark sector DSE 
technique, by setting the quark mass to large values in 
the existing formalism and computer codes, simply will not work.  This is 
because the interesting physical phenomena are driven by non-perturbative 
contributions to the heavy quark propagator in the vicinity of the bare 
fermion mass pole. These contributions are of order $1/m_Q$ compared 
with the bare fermion propagator.   
Earlier studies of the heavy fermion limit of another confining theory, 
QED3 \cite{A96}, demonstrate categorically that 
for fermions whose mass is greater than the typical scale of the interaction 
of the theory, this information cannot be accurately obtained by an 
extrapolation of the spacelike solution of the fermion propagator 
produced by the usual DSE technique, and a new formalism must be developed.  

We begin with the general form for the inverse of the dressed quark 
propagator consistent with Lorentz and CPT covariance: 
\begin{equation}
S^{-1}(p) = i \gamma\cdot p A(p^2) + B(p^2),  \label{genprop}
\end{equation}
where $A$ and $B$ are scalar functions.  We use a Euclidean metric in which 
timelike 4-vectors satisfy $p^2 < 0$.  Within the DSE technique the full 
quark Dyson-Schwinger equation is typically truncated to a manageable 
approximation.  For the purposes of illustrating the extension to the 
heavy quark sector, we consider the simplest viable approximation 
to the quark DSE,
\begin{equation}
S^{-1}(p) - i\gamma\cdot p - m_Q = 
 \frac{4}{3} \int \frac{d^4q}{(2\pi)^4} 
                \gamma_\mu S(q) \gamma_\mu \Delta(p - q),  \label{SDEqn}
\end{equation}
in which the quark--gluon vertex has been replaced by a bare vertex, and 
the dressed gluon propagator is modelled by a representation 
$g^2D_{\mu \nu}(k) = \delta_{\mu \nu} \Delta(k^2)$ in a Feynman-like gauge.  
Our formalism for calculating model heavy quark propagators can be extended 
without trouble to include more realistic vertex ans\"{a}tze (to allow 
for gauge covariance \cite{RW94}) and to more realistic model gluon 
propagators (see, for instance, \cite{FR96}).  

In the limit of heavy quark masses, $m_Q \rightarrow \infty$, the dressed 
quark propagator is dominated by the bare form $S^{-1}_{\rm bare} = 
i\gamma\cdot p + m_Q$.  However it is important to isolate from the full 
propagator order $1/m_Q$ corrections to the bare propagator which drive 
confining and remnant chiral symmetry breaking effects.  To this end we 
set 
\begin{equation}
A(p^2) = 1 + \frac{\Sigma_A(k)}{m_Q},\hspace{5 mm}
B(p^2) = m_Q\left(1 + \frac{\Sigma_B(k)}{m_Q}\right),    \label{SABDEF}
\end{equation}
where, in the spirit of HQET, we have introduced a new momentum variable 
\begin{equation}
p_\mu = im_Q v_\mu + k_\mu,     \label{kdef}
\end{equation}
with $v_\mu$ a constant unit 4-vector.  

The functions $\Sigma_A$ and $\Sigma_B$ can be expanded as a series in 
$1/m_Q$, and the DSE solved order by order as a set of coupled integral 
equations.  In the appendix we derive the set of four coupled equations 
required to solve the heavy quark propagator to $O(1/m_Q)$.  To leading 
order, we find that $\Sigma_A$ and $\Sigma_B$ can be written as functions 
of the single variable $k\cdot v$, and the quark propagator becomes 
\begin{equation}
S(p) = \frac{1 + \!\not \! v}{2} \frac{1}
       {ik\cdot v + \Sigma_B(k\cdot v) - \Sigma_A(k\cdot v)} 
         + O\left(m_Q^{-1}\right).  \label{heavy}
\end{equation}
Substituting into Eq.~(\ref{SDEqn}), changing the variable of integration 
to $k'$ defined by $q_\mu = imv_\mu + k'_\mu$, and projecting out 
coefficients of $\not \! v$, $\not \! k$ and $I$ gives a set of integral 
equations which can be summarised in the equation (see Eqs.~(\ref{a7}) 
and (\ref{a8}))
\begin{equation}
\Sigma(k\cdot v) = \frac{4}{3} \int \frac{d^4k'}{(2\pi)^4} \Delta(k - k')
    \frac{1}{i k'\cdot v + \Sigma(k'\cdot v) } \,  , \label{sga}
\end{equation}
where 
\begin{equation}
\Sigma(k\cdot v) = \Sigma_B(k\cdot v) - \Sigma_A(k\cdot v).  \label{sgb}
\end{equation} 
Note that to leading order in $1/m_Q$ the heavy quark self energy is 
specified by a single, complex valued, scalar function.  
This equation is numerically tractable once the function $\Delta$ is 
specified,  and provides the required information regarding fermion self 
dressing in the vicinity of the bare fermion mass pole, $k_\mu \approx 0$,
in the limit $m_Q \rightarrow \infty$.  

Note also that the change of integration variable from $q$ to $k'$ involves 
an assumption that the propagator, and hence the functions 
$A(q^2)/[q^2A(q^2)^2 + B(q^2)^2]$ and $B(q^2)/[q^2A(q^2)^2 + B(q^2)^2]$, 
be analytic over the region 
\begin{equation}
{\rm Re}\,(q^2) > -m_Q^2 + \frac{\left({\rm Im}\,(q^2)\right)^2}{4m_Q^2}.  
\end{equation}
In the limit $m_Q\rightarrow \infty$ this translates to the requirement 
that $1/[ik\cdot v + \Sigma(k\cdot v)]$ be analytic for 
${\rm Im}(k\cdot v) < 0$.  

\setcounter{equation}{0}
\section{The Bethe-Salpeter equation for heavy-light mesons.} 

The combination of bare vertex DSE and ladder approximation Bethe-Salpeter
equation produces well the spectrum of light quark 
mesons \cite{BSp1,BSpapers}.  
Here we extend the formalism to mesons containing one heavy quark 
and one light antiquark in the limit $m_Q \rightarrow \infty$.  
We note that our work differs in a fundamental way from that of 
Villate {\it et al.}\cite{V93} who use an instantaneous approximation 
to study heavy quarkonium states via the combination of DSE and BSE.  
Because the light quark present in our case must be treated in a fully 
relativistic fashion, the instantaneous approximation is not applicable.  

The homogeneous ladder approximation to the meson Bethe-Salpeter
equation in a Feynman-like gauge is,
\begin{equation}
\Gamma(p,P) = - \frac{4}{3}\int \, \frac{d^4q}{(2\pi)^4}\, 
          \Delta(p - q) \gamma_\mu S_q(q - \xi P)
            \Gamma(q,P) S_Q(q + (1 - \xi) P) \gamma_\mu. \label{bse}
\end{equation}
Here $S_q(p)$ is the full light quark propagator, $S_Q(p)$ the 
full heavy quark propagator, and $\delta_{\mu\nu} \Delta(p - q)$ the 
same effective gluon propagator as that employed in the DSE.  
The Bethe-Salpeter amplitude $\Gamma(q,P)$ is defined so that external
outgoing and incoming quark lines carry momenta $q + (1 - \xi)P$ and
$q - \xi P$ respectively.  

For the light quark, we write the propagator as 
\begin{eqnarray}
S_q(p) & = & -i \not\! p \sigma_V(p^2)+\sigma_S(p^2) \nonumber \\
     & = & \frac{1}{ i \not\! p A_q(p^2)+B_q(p^2) }.   \label{ltprop}
\end{eqnarray}
The vector and scalar parts $\sigma_V$ and $\sigma_S$ are
related to the functions $A_q$ and $B_q$ simply by dividing
by the quantity $p^2 A_q^2(p^2) + B_q^2(p^2)$.  Using Eq.~(\ref{SDEqn}), 
with the quark mass replaced by the light current quark mass $m_q$, 
$A_q$ and $B_q$ are given by the integral equations 
\begin{eqnarray}
p^2\left(A_q(p^2) - 1\right) = \frac{8}{3} \int \, \frac{d^4q}{(2\pi)^4}\, 
  p\cdot q
    \frac{A_q(q^2)}{q^2 A_q^2(q^2) + B_q^2(q^2)} \Delta(p - q),  \label{aeqn}
\end{eqnarray}
\begin{eqnarray}
B_q(p^2) - m_q = \frac{16}{3} \int \, \frac{d^4q}{(2\pi)^4}\, 
    \frac{B_q(q^2)}
      {q^2 A_q^2(q^2) + B_q^2(q^2)} \Delta(p - q).     \label{beqn}
\end{eqnarray}
For the heavy quark we use the leading order heavy quark propagator 
Eq.~(\ref{heavy}), 
\begin{equation}
S_Q(im_Q v_\mu + k_\mu) = \frac{1 + \!\not \! v}{2} \frac{1}
       {ik\cdot v + \Sigma(k\cdot v)} \, \label{heavy1}
\end{equation}
where $\Sigma(k\cdot v)$ is obtained from Eq.~(\ref{sga}).  
 
We choose to work in the rest frame of the meson and set 
\begin{equation}
P_\mu = \left({\bf 0},i(m_Q + \delta)\right)   \label{rf}
\end{equation}
where $\delta$ is the contribution to the meson mass from the ``brown muck''
\cite{N94} of light quarks and glue surrounding the heavy quark.  Our 
aim is to obtain the BSE in the limit $m_Q \rightarrow \infty$ in a form 
containing $\delta$, but independent of $m_Q$.  It is convenient to 
reparameterise the momentum partitioning in Eq.~(\ref{bse}) via 
\begin{equation}
\xi = \frac{\eta \delta}{m_Q + \delta}.  \label{etadef}
\end{equation}
Since the light quark Eqs.~(\ref{aeqn}) and (\ref{beqn}) 
are solved in the first instance for real, spacelike momenta, we see that 
the choice $\eta = 0$ has the advantage that no numerical extrapolation 
of $S_q$ to complex momenta is required.  On the other hand, 
using the freedom inherent in $v_\mu$ to set $v_\mu =  ({\bf 0},1)$, 
we see the choice $\eta = 1$ has the advantage that no extrapolation of 
the numerical solution of the heavy quark self energy $\Sigma(k_4)$ is 
required away from the real $k_4$ axis.  In practice, one expects that 
the optimum choice of $\eta$ requiring minimal numerical 
extrapolation of either propagator lies somewhere between these values.  

Before deriving integral equations for scalar and vector heavy-light mesons, 
it is convenient make some definitions.  Define the projection operators 
\begin{equation}
\Lambda_\pm = (1 \pm \gamma_4)/2. 
\end{equation}
The light quark propagator can then be written as 
\begin{equation}
S_q(q - \xi P) =  
     S_+\Lambda_+  + S_-\Lambda_- + S_\perp \! \not \! \hat{\bf q}, 
                       \label{lightq}
\end{equation}
where 
\begin{equation}
\not \! \hat{\bf q} = (q_1 \gamma_1 + q_2 \gamma_2 + q_3 \gamma_3)/
                     \left|{\bf q}\right|, 
\end{equation}
with $\left|{\bf q}\right| = \sqrt{q_1^2 + q_2^2 + q_3^2}$ and 
\begin{equation}
S_+(q_4,\left|{\bf q}\right|) = 
    \sigma_S(s) - (\eta\delta + iq_4)\sigma_V(s),  
                                                   \label{splusdef}
\end{equation}
\begin{equation}
S_-(q_4,\left|{\bf q}\right|) = 
    \sigma_S(s) + (\eta\delta + iq_4)\sigma_V(s), 
                                                   \label{sminusdef}
\end{equation}
\begin{equation}
S_\perp(q_4,\left|{\bf q}\right|) = - i\left|{\bf q}\right| \sigma_V(s), 
                                                 \label{sperpdef}
\end{equation}
where
\begin{equation}
s = \left|{\bf q}\right|^2 + q_4^2 - \eta^2 \delta^2 - 2i\eta\delta q_4.  
                       \label{ess}
\end{equation}
With $v_\mu = ({\bf 0},1)$, the heavy quark propagator is, from 
Eq.~(\ref{heavy1}), 
\begin{equation}
S_Q(q + (1 - \xi) P) = \frac{\Lambda_+}
  {iq_4 - (1 - \eta)\delta + \Sigma\left(q_4 + i(1 - \eta)\delta\right)} \,.
                                             \label{heavyq}
\end{equation}
We see that, by judicious choice of momentum partitioning, the quark 
propagators, and hence the BSE, are independent of $m_Q$ in the 
$m_Q \rightarrow \infty$ limit.  

\setcounter{equation}{0}
\section{Scalar and Pseudoscalar mesons.}

The general forms of the pseudoscalar and scalar Bethe-Salpeter amplitudes 
are given by \cite{LS69}
\begin{equation}
\Gamma^{\rm pseud}(q,P) = 
\left(f_1 + f_2 \not \! P + f_3 \not \! q + f_4 [\not \! q , \not \! P]
      \right) \gamma_5,    \label{genps}
\end{equation}
\begin{equation}
\Gamma^{\rm scalar}(q,P) = 
f_1 + f_2 \not \! P + f_3 \not \! q + f_4 [\not \! q , \not \! P],
          \label{gensc}
\end{equation}
where $f_1$ to $f_4$ are functions of $q^2$, $P^2$ and $q\cdot P$.  
With the $P_\mu$ chosen as Eq.~(\ref{rf}), these can equally well be 
written in terms of the projection operators $\Lambda_\pm$ as 
\begin{equation}
\Gamma^{\rm pseud}(q) = 
\left(f_+ \Lambda_+ + f_- \Lambda_- + g_+ \! \not \! \hat{\bf q} \Lambda_+
                   + g_- \! \not \! \hat{\bf q} \Lambda_-
      \right) \gamma_5,    \label{genps2}
\end{equation}
\begin{equation}
\Gamma^{\rm scalar}(q) = 
f_+ \Lambda_+ + f_- \Lambda_- - g_+ \! \not \! \hat{\bf q} \Lambda_+
                   - g_- \! \not \! \hat{\bf q} \Lambda_-,    \label{gensc2}
\end{equation}
where $f_\pm$ and $g_\pm$ are functions of $q_4$ and $\left|{\bf q}\right|$.

By substituting Eqs.~(\ref{lightq}) and (\ref{heavyq}) for the quark 
propagators into Eq.~(\ref{bse}) and making use of the identities 
\begin{equation}
\gamma_\mu \Lambda_\pm \gamma_\mu = \Lambda_\pm + 3\Lambda_\mp, \hspace{5 mm}
\gamma_\mu \! \not \! \hat{\bf q} \Lambda_\pm \gamma_\mu 
                    = - \! \not \! \hat{\bf q}. 
\end{equation}
a set of coupled integral equations can be projected out.  For the 
pseudoscalar meson we obtain 
\begin{equation}
f_-(p) = \frac{4}{3} \int \frac{d^4q}{(2\pi)^4} \Delta(p - q)
  \frac{S_- f_-(q) + S_\perp g_-(q)}
    {iq_4 - (1 - \eta)\delta + \Sigma\left(q_4 + i(1 - \eta)\delta\right)},
                           \label{pseudf}
\end{equation}
\begin{equation}
f_+(p) = 3f_-(p)
\end{equation}
\begin{equation}
g_-(p) = -\frac{4}{3} \int \frac{d^4q}{(2\pi)^4} 
       \hat{\bf p}.\hat{\bf q} \Delta(p - q)
  \frac{S_+ g_-(q) + S_\perp f_-(q)}
    {iq_4 - (1 - \eta)\delta + \Sigma\left(q_4 + i(1 - \eta)\delta\right)},
                            \label{pseudg}
\end{equation}
\begin{equation}
g_+(p) = g_-(p).
\end{equation}
The scalar equations can be obtained from these by making the replacements 
$\sigma_S \rightarrow -\sigma_S$ in Eqs.~(\ref{splusdef}) and 
(\ref{sminusdef}). 

\setcounter{equation}{0}
\section{Vector and axial vector mesons.} 

Llewellyn-Smith \cite{LS69} gives the general form 
of the $J^P = 1^-$ BS amplitude as 
\begin{eqnarray}
\Gamma_\mu(q,P) & = & T_\mu\left(A + q\cdot P B \not\! q + C \not\! P + 
	     D[\not\! q ,\not\! P]\right)\nonumber \\
& & -Q_\mu(q\cdot P B + 2 \not \! P D ) \nonumber \\
& & +Q_\mu\left(E + F \not\! q + q\cdot P G \not\! P + 
	     H[\not\! q ,\not\! P]\right) , \label{LSgen}
\end{eqnarray}
where
\begin{equation}
T_\mu = \left(\delta_{\mu \nu} - \frac{P_\mu P_\nu}{P^2} \right)\gamma_\nu,
\hspace{5 mm}
Q_\mu = \left(\delta_{\mu \nu} - \frac{P_\mu P_\nu}{P^2} \right)q_\nu,
\end{equation}
and $A$ to $H$ are functions of $q^2$, $P^2$ and $q\cdot P$.  
The second line of Eq.~(\ref{LSgen}) serves the purpose of ensuring that 
$A$ to $H$ are even functions of $q\cdot P$ if the fermion is odd under 
charge conjugation.  Since we are interested in $q$-$\overline{Q}$
mesons, charge conjugation is not important, and this line can be 
ignored.  

Working in the vector meson rest frame, Eq.~(\ref{rf}), the general form 
of the amplitude can be recast in the form $\Gamma_\mu = (\vec{\Gamma},0)$ 
with
\begin{equation}
\vec{\Gamma}(q) = \vec{\Gamma}^+(q) \Lambda^+ + \vec{\Gamma}^-(q) \Lambda^-,
		 \label{genvec}
\end{equation} 
where 
\begin{equation}
\vec{\Gamma}^\pm(q) = \vec{\gamma}\left(f_\pm + \not \! 
  \hat{\bf q} g_\pm \right) 
	 + \hat{\bf q}\left(h_\pm + \not \! \hat{\bf q} k_\pm \right), 
\end{equation}
with $f_\pm$, $g_\pm$, $h_\pm$ and $k_\pm$ functions of $q_4$ and 
$\left|{\bf q}\right|$, and $\vec{\gamma} = (\gamma_1,\gamma_2,\gamma_3)$ 
and $\hat{\bf q} = (q_1,q_2,q_3)/\left|{\bf q}\right|$.  Using the identity 
$\vec{\gamma} \!\not \!\! \hat{\bf q} = \hat{\bf q} + 
            \frac{1}{2}[\vec{\gamma},\not \!\! \hat{\bf q}]$, 
it is convenient to absorb part of the $g_\pm$ terms into a redefined  
$h_\pm$ and write instead 
\begin{equation}
\Gamma^\pm_i(q) = \left(\delta_{ij}f_\pm + \hat{q}_i \hat{q}_j k_\pm \right)
\gamma_j + \frac{1}{2} g_\pm[\gamma_i,\not \! \hat{\bf q}] + h_\pm \hat{q}_i. 
                                   \label{vec2}
\end{equation}
The following relations are then useful for extracting the scalar coefficient 
functions from $\Gamma^\pm_i$: 
\begin{equation}
f_\pm(p) = \frac{1}{8}\left(\delta_{ij} - \hat{p}_i \hat{p}_j \right)
	 {\rm tr}\left(\gamma_k \Gamma^\pm_i(p)\right), 
			       \label{projf} 
\end{equation}
\begin{equation}
g_\pm(p) = -\frac{1}{16}
{\rm tr}\left([\gamma_i, \not \! \hat{\bf p}]\Gamma^\pm_i(p)\right), 
			       \label{projg} 
\end{equation}
\begin{equation}
h_\pm(p) = \frac{1}{4} {\rm tr}\left(\hat{p}_i\Gamma^\pm_i(p)\right), 
			       \label{projh} 
\end{equation}
\begin{equation}
k_\pm(p) = \frac{1}{8}\left(3\hat{p}_i \hat{p}_j - \delta_{ij} \right)
	 {\rm tr}\left(\gamma_k \Gamma^\pm_i(p)\right). 
			       \label{projk}  
\end{equation}

Substituting Eq.~(\ref{vec2}) into the BSE Eq.(\ref{bse}) together 
with the propagators Eqs. (\ref{lightq}) and (\ref{heavyq}), and making 
use of the identities 
\begin{equation}
\gamma_\mu \Lambda_\pm \gamma_\mu = \Lambda_\pm + 3\Lambda_\mp, 
\hspace{5 mm}
\gamma_\mu \gamma_i \Lambda_\pm \gamma_\mu = 
-\gamma_i(\Lambda_\pm + \Lambda_\mp), 
\end{equation}
\begin{equation}
\gamma_\mu [\gamma_i,\not \! \hat{\bf q}] \Lambda_\pm \gamma_\mu = 
[\gamma_i,\not \! \hat{\bf q}](\Lambda_\pm - \Lambda_\mp), 
\end{equation}
we obtain the following set of coupled integral 
equations: 
\begin{eqnarray}
\lefteqn{f_+(p) = -\frac{4}{3} \int \, \frac{d^4q}{(2\pi)^4} 
               \Delta(p - q)\times}\nonumber \\
& &  \!\!\!\!\!\left[ 
  \left(-S_- f_+(q) + S_\perp g_+(q)\right) - 
   \frac{1}{2}\left(S_-k_+ + S_\perp (g_+ + h_+)\right) 
                \left(1 - (\hat{\bf p}.\hat{\bf q})^2 \right)\right]\times
                                                    \nonumber \\ 
& &   \frac{1}
    {iq_4 - (1 - \eta)\delta + \Sigma\left(q_4 + i(1 - \eta)\delta\right)},
\end{eqnarray}
\begin{equation}
f_-(p) = f_+(p),
\end{equation}
\begin{eqnarray}
\lefteqn{g_+(p) = -\frac{4}{3} \int \, \frac{d^4q}{(2\pi)^4} \, 
         \Delta(p - q)\times}\nonumber \\
& &  
  \left(S_+ g_+(q) - S_\perp f_+(q)\right) \hat{\bf p}.\hat{\bf q}
  \frac{1}
    {iq_4 - (1 - \eta)\delta + \Sigma\left(q_4 + i(1 - \eta)\delta\right)},
\end{eqnarray}
\begin{equation}
g_-(p) = -g_+(p),
\end{equation}
\begin{eqnarray}
\lefteqn{h_+(p) = -\frac{4}{3} \int \, \frac{d^4q}{(2\pi)^4} \, 
                      \Delta(p - q)\times}\nonumber \\
& &  \!\!\! 
  \left(S_+ h_+(q) + S_\perp (f_+(q) + k_+(q)\right) \hat{\bf p}.\hat{\bf q}
  \frac{1}
    {iq_4 - (1 - \eta)\delta + \Sigma\left(q_4 + i(1 - \eta)\delta\right)},
\end{eqnarray}
\begin{equation}
h_-(p) = 3h_+(p),
\end{equation}
\begin{eqnarray}
\lefteqn{k_+(p) = -\frac{4}{3} \int \, \frac{d^4q}{(2\pi)^4} \,
                     \Delta(p - q)\times}  \\
& &  \!\!\!\!\!\!\!\!\!\! 
  \frac{1}{2}\left[S_-k_+(q) + S_\perp (g_+(q) + h_+(q))\right]
                \left(1 - 3(\hat{\bf p}.\hat{\bf q})^2 \right)
  \frac{1}
    {iq_4 - (1 - \eta)\delta + \Sigma\left(q_4 + i(1 - \eta)\delta\right)},
                              \nonumber
\end{eqnarray}
\begin{equation}
k_-(p) = k_+(p).
\end{equation}
These equations can be simplified greatly by setting 
\begin{equation}
\tilde{f} = 3f_+ + k_+,
\end{equation}
\begin{equation}
\tilde{g} = h_+ - 2g_+,
\end{equation}
\begin{equation}
\tilde{h} = g_+ + h_+,
\end{equation}
\begin{equation}
\tilde{k} = k_+.
\end{equation}
This gives 
\begin{equation}
\tilde{f}(p) = \frac{4}{3} \int \, \frac{d^4q}{(2\pi)^4} \, 
                      \Delta(p - q)
  \frac{S_-(q) \tilde{f}(q) + S_\perp(q) \tilde{g}(q)}
    {iq_4 - (1 - \eta)\delta + \Sigma\left(q_4 + i(1 - \eta)\delta\right)},
                                         \label{fvec}
\end{equation}
\begin{equation}
\tilde{g}(p) = -\frac{4}{3} \int \, \frac{d^4q}{(2\pi)^4} \, 
                    \hat{\bf p}.\hat{\bf q} \Delta(p - q)
  \frac{S_+(q) \tilde{g}(q) + S_\perp(q) \tilde{f}(q)}
    {iq_4 - (1 - \eta)\delta + \Sigma\left(q_4 + i(1 - \eta)\delta\right)},
                                    \label{gvec}
\end{equation}
\begin{equation}
\tilde{h}(p) = -\frac{4}{3} \int \, \frac{d^4q}{(2\pi)^4} \, 
              \hat{\bf p}.\hat{\bf q} \Delta(p - q)
  \frac{S_+(q) \tilde{h}(q) + S_\perp(q) \tilde{k}(q)}
    {iq_4 - (1 - \eta)\delta + \Sigma\left(q_4 + i(1 - \eta)\delta\right)},
                                   \label{hvec}
\end{equation}
\begin{equation}
\tilde{k}(p) = -\frac{2}{3} \int \, \frac{d^4q}{(2\pi)^4} \, 
       \left(1 - 3(\hat{\bf p}.\hat{\bf q})^2\right) \Delta(p - q)
  \frac{S_-(q)\tilde{k}(q) + S_\perp(q) \tilde{h}(q)}
    {iq_4 - (1 - \eta)\delta + \Sigma\left(q_4 + i(1 - \eta)\delta\right)}.
                     \label{kvec}
\end{equation}
Eqs.~(\ref{fvec}) and (\ref{gvec}) are equivalent to the scalar equations 
Eqs.~(\ref{pseudf}) and (\ref{pseudg}). They represent the  
$1^-$ meson whose light quark spin content is $j = \frac{1}{2}$, which, 
as predicted by HQET, is degenerate with the 
scalar $0^-$ meson.  Eqs.~(\ref{hvec}) and (\ref{kvec}) represent the 
$1^-$ meson whose light quark spin content is $j = \frac{3}{2}$.  

The $1^+$ meson Bethe-Salpeter amplitude is obtained by multiplying the 
general form Eq.~(\ref{genvec}) through by $\gamma_5$. The corresponding set 
of integral equations can be obtained from those above by making the 
replacement $\sigma_S \rightarrow -\sigma_S$ in Eqs.~(\ref{splusdef}) 
and (\ref{sminusdef}).  

\setcounter{equation}{0}
\section{Numerical Calculations.}

In order to explore the potential of our formalism we consider here the 
simple gaussian gluon propagator ansatz 
\begin{equation}
\Delta(p - q) = \frac{3}{16} (2\pi)^4 \frac{\mu^2}{\alpha^2 \pi^2}
              e^{-(p - q)^2/\alpha}. \label{gss}
\end{equation}
This model propagator is simple enough to render the angular integrals 
in both the SDE and the BSE analytically tractable, but has the disadvantage 
that it underestimates the gluon interaction strength in the ultraviolet.  
However, the evidence from light meson applications of the SDE technique 
is that it is the infrared behaviour of the gluon propagator which most 
influences the low energy physics.  Furthermore, the large mass expansion 
makes sense for non-perturbative interactions, but not for hard gluons.  
By using a gaussian model gluon propagator we provide an ultraviolet 
cutoff within a treatment which does not admit hard gluon contributions.  
We therefore regard this model propagator 
as an acceptable starting point.  For small $\alpha$ Eq.~(\ref{gss}) 
reduces to the infrared dominant model 
\begin{equation}
\lim_{\alpha\rightarrow 0}\Delta(p - q) 
  = \frac{3}{16} (2\pi)^4 \mu^2 \delta^4(p - q), \label{dlta}
\end{equation}
which has frequently been employed in exploratory applications of the DSE 
technique to light quarks~\cite{Deltafn,B96}.  

Substituting Eq.~(\ref{gss}) into the heavy quark DSE Eq.~(\ref{sga}) 
with $v_\mu = ({\bf 0},1)$ gives 
\begin{equation}
\Sigma(k_4) = \frac{\mu^2}{4(\alpha\pi)^{1/2}} \int_0^\infty dk_4'
  \left[\frac{e^{-(k_4 - k_4')^2/\alpha}}{ ik_4' + \Sigma(k_4')  } +
       \frac{e^{-(k_4 + k_4')^2/\alpha}}{-ik_4' + \Sigma(k_4')^*}\right].
                            \label{sigie}
\end{equation}
Here we have assumed $\Sigma(-k_4) = \Sigma(k_4)^*$, this being 
a property of the obvious extension to negative $k_4$ of the solution 
\begin{equation}
\lim_{\alpha\rightarrow 0}\Sigma(k_4) =
\left \{ \begin{array}{lrl}
\frac{1}{2} \left[-ik_4 + \sqrt{\mu^2 - k_4^2} \right]
                                  & \mbox{if } & 0\le k_4 < \mu , \\
                                  &            &            \\
\frac{i}{2} \left[- k_4 + \sqrt{k_4^2 - \mu^2} \right]
                                  & \mbox{if } & k_4\ge \mu .
                   \end{array}   \right.   \label{delsol}
\end{equation}
to the algebraic equation arising from the infrared dominant propagator 
Eq.~(\ref{dlta}).  
Substituting the gaussian propagator Eq.~(\ref{gss}) into the light quark 
DSE Eqs. (\ref{aeqn}) and (\ref{beqn}) gives
\begin{equation}
A_q(p^2) - 1 = \frac{\mu^2}{\alpha p^2} \int_0^\infty
     q^3 \, dq \, \frac{A_q(q^2)}{q^2A_q(q^2)^2 + B_q(q^2)^2}
                e^{-(p^2 + q^2)/\alpha} I_2\left(\frac{2pq}{\alpha}\right),
                              \label{Aeq}
\end{equation}
\begin{equation}
B_q(p^2) - m_q = \frac{2\mu^2}{\alpha p} \int_0^\infty
     q^2 \, dq \, \frac{B_q(q^2)}{q^2A_q(q^2)^2 + B_q(q^2)^2}
                e^{-(p^2 + q^2)/\alpha} I_1\left(\frac{2pq}{\alpha}\right),
                              \label{Beq}
\end{equation}
where $I_1$ and $I_2$ are modified Bessel functions.  

Here we shall carry out an initial study of the viability of the model 
gaussian gluon propagator in the light quark chiral limit 
$m_q \rightarrow 0$, and heavy quark limit $m_Q \rightarrow \infty$.  
In this limit the parameter $\alpha$ in Eq.~(\ref{gss}) can be scaled to 
unity in all equations by making the replacements 
\begin{equation}
\mu = \sqrt{\alpha}\hat{\mu}, \hspace{5 mm}
\delta = \sqrt{\alpha}\hat{\delta},    \label{scale1}
\end{equation}
\begin{equation}
\Sigma(k_4) = \sqrt{\alpha}\hat{\Sigma}(\hat{k}_4), \hspace{5 mm}
A(p^2) = \hat{A}(\hat{p}^2), \hspace{5 mm}
B(p^2) = \sqrt{\alpha}\hat{B}(\hat{p}^2), \label{scale2} 
\end{equation}
where 
\begin{equation}
\hat{k}_4 = k_4/\sqrt{\alpha}, \hspace{5 mm}
\hat{p} = p/\sqrt{\alpha}.      \label{scale3}
\end{equation}
It is then sufficient to explore the parameter space of the single parameter 
$\hat{\mu}$.  

We have solved Eqs. (\ref{Aeq}) and (\ref{Beq}) by numerical iteration 
for real, spacelike $\hat{p}^2$ for a range of values of $\hat{\mu}$. 
To extrapolate the light quark propagator from the spacelike axis into the 
complex $p^2$-plane, the right hand sides of Eqs.~(\ref{aeqn}) 
and (\ref{beqn}) were integrated along the real spacelike $q^2$-axis using 
the converged solutions together with complex values of $\hat{p}^2$.  
We find that for $\hat{\mu} < 3$, the light quark propagator has a timelike 
pole, and that for larger values of $\hat{\mu}$ the pole moves away from the 
real axis into the complex plane.  Table~\ref{tabpole} lists the position 
$\hat{p}^2 = \hat{s}^{\rm pole}$ of the pole as a function of $\hat{\mu}$.  

To obtain a numerical solution to Eq.~(\ref{sigie}) for real $\hat{k}_4$ we 
iterated from Eq.~(\ref{delsol}).  The numerical solution for 
$\hat{\mu} = 3.5$ is shown in Fig.~1.  The function $\hat{\Sigma}$ is 
extended to complex values of $\hat{k}_4$ 
by integrating the right hand side of Eq.~(\ref{sigie}) along the real 
$k_4'$-axis.  We find numerically that the requirement stated at the end 
of Section~III, viz. that the heavy quark propagator be free of 
singularities for ${\rm Im}\,k_4 < 0$, is satisfied over the range of 
$\hat{\mu}$ listed in Table~\ref{tabpole}.  However, we also find a succession 
poles which migrate inwards from $i\infty$ along the positive imaginary 
$k_4$-axis before veering away from the axis as $\hat{\mu}$ is increased.  
The position of these poles is plotted in Fig.~2, and the position of 
the pole closest to the real $k_4$-axis is listed in Table~\ref{tabpole}.  
We note also that, as $\hat{\mu} \rightarrow \infty$ corresponding to the 
infrared dominant limit Eq.~(\ref{dlta}), the poles move further from the 
origin.  This is consistent with the limiting solution Eq.~(\ref{delsol}) 
which is free from poles except at $k_4 = \infty$.  

If the quarks are to be a confined particles, we are restricted to 
values of $\mu$ for which the light quark propagator $S_q(p)$ has no 
pole on the real timelike $p^2$-axis, and for which the heavy quark 
propagator $1/\left(ik_4 + \Sigma(k_4)\right)$ has no poles on the 
imaginary $k_4$-axis.  While this requirement can be fulfilled by our 
light quark propagator solutions, unfortunately we find the heavy quark 
propagator is not totally free of timelike poles for any value of $\hat{\mu}$ 
over the range considered.  It is reasonable to suggest that this fault 
could be rectified by improving 
the approximations employed DSE (bare vertex and gaussian 
gluon propagtor).  In fact, in ref.~\cite{BRW92}, it is shown that if 
the infrared dominant gluon propagator Eq.~(\ref{dlta}) is used in 
conjunction with a vertex ansatz satisfying the Ward-Takahashi 
identity, one obtains a quark propagator which is an entire function 
of $p^2$ irrespective of the quark mass.  It is therefore reasonable 
to assume that the timelike poles in the heavy quark propagator are an 
artifact of the model.  

Poles which occur away from the timelike axis are, in principle, no 
impediment to confinement.  However, solution of the BSE requires 
knowledge of the quark propagators over a region of the complex plane.  
From Eqs.~(\ref{lightq}) to (\ref{heavyq}), we see that the light quark 
propagator $S_q(p)$ is sampled by the BSE over the region 
\begin{equation}
{\rm Re}\,s > \frac{({\rm Im}\,s)^2}{4\eta^2 \delta^2} - \eta^2 \delta^2,  
                                      \label{rsts}
\end{equation}  
where $s = p^2$, and the heavy quark propagator is sampled along the line 
\begin{equation}
{\rm Im}\,k_4 = (1 - \eta)\delta,   \label{rstk}
\end{equation}
where $\delta$ is the ``brown muck'' contribution to the meson mass, and 
$\eta$ a momentum partitioning parameter.  
Assuming there are no compensating zeros in the BS amplitudes, the integral 
in the BSE is divergent if poles are encounted in either of these 
two regions.  Writing $\hat{s}^{\rm pole} = X + iY$ and 
${\rm Im}\,\hat{k}_4^{\rm pole} = K$, 
it is straightforward to show that the model then fails if 
\begin{equation}
\hat{\delta} > \hat{\delta}_{\rm max} = 
     K + \left[\frac{\sqrt{X^2 + Y^2} - X}{2}\right]^{1/2}.  \label{dmax}
\end{equation}
To approach the maximum allowed value of $\hat{\delta}$ in numerical 
calculations, it is necessary to choose the momentum partitioning 
$\eta$ to take its optimum value 
\begin{equation}
\eta_{\rm opt} = \frac{\left[\sqrt{X^2 + Y^2} - X \right]^{1/2}}
                {\sqrt{2}K + \left[\sqrt{X^2 + Y^2} - X \right]^{1/2}}.  
\end{equation}
Values of $\hat{\delta}_{\rm max}$ and $\eta_{\rm opt}$ are plotted in 
Table~\ref{tabpole}.  

Solution of the BS equations involves the iteration of the sets of 
coupled linear integral equations obtained in Sections V and VI.  For 
the $J^P = 0^-$ meson, for instance, from Eqs.~(\ref{pseudf}) and 
(\ref{pseudg}) and the definitions (\ref{splusdef}) to (\ref{sperpdef})
we can assume without loss of generality that 
\begin{equation}
f_-(\left|{\bf q}\right|,-q_4^*) =
      f_-(\left|{\bf q}\right|,q_4)^*, \hspace{5 mm}
g_-(\left|{\bf q}\right|,-q_4^*) =
     -g_-(\left|{\bf q}\right|,q_4)^*.
\end{equation}
Executing the spatial angular interations analytically then gives a 
set of equations of the form 
\begin{equation}
\vec{f}(\left|{\bf p}\right|,p_4) = \int_0^\infty dq_4 
     \int_0^\infty d\left|{\bf q}\right|\, 
  K(\left|{\bf p}\right|,p_4;\left|{\bf q}\right|,q_4;\delta)
            \vec{f}(\left|{\bf q}\right|,q_4), 
\end{equation}
where $\vec{f} = 
({\rm Re}\,f_-,{\rm Im}\,f_-,{\rm Re}\,g_-,{\rm Im}\,g_-)^{\rm T}$ 
and the function $K$ is a $4\times 4$ matrix.  One solves this as an 
eigenvalue equation of the form 
\begin{equation}
\int dq\, K(p,q;\delta) \vec{f}(q) = \Lambda(\delta) \vec{f}(p),
                    \label{lineq}  
\end{equation}
for a range of test values of $\delta$ until an eigenvalue 
$\Lambda(\delta) = 1$ is obtained.  

The numerical solution to Eq.~(\ref{lineq}) and its analogues for the 
negative parity states considered in Sections V and VI was carried out 
by linearly interpolating the functions contained in $\vec{f}$ on 
a $10\times 10$ 
grid and then linearly interpolating the integrand on a $50\times 50$ 
grid.  A cutoff of $\left|\hat{\bf q}\right|,\hat{q}_4 < 10$ in the 
dimensionless units defined by Eq.~(\ref{scale3}) was found to be adequate.  

The calculations were done for the dimensionless parameter values 
$\hat{\mu} = 3.5$, 5.0 and 6.0.  These values are chosen to give a broad 
range of $\hat{\mu}$ while ensuring that 
the light propagator is confining (i.e. free from poles on the real 
timelike axis).  Poles on the imaginary $k_4$-axis rendering the 
heavy the heavy quark propagator non-confining were sufficiently far away 
from the region of integration of the BSE to be considered irrelevant.  
However, one must also consider poles elsewhere in the the complex plane 
of both the light and heavy quark propagators, namely those listed in 
Table~\ref{tabpole}, and the consequent restriction Eq.~(\ref{dmax}) on 
numerically accessible values of $\hat{\delta}$.  To allow for as large a 
range of accessible $\hat{\delta}$ as possible, the momentum partitioning 
$\eta_{\rm opt}$ from Table~\ref{tabpole} was used.  

In Fig.~3 we plot $\Lambda(\hat{\delta})$ for the degenerate $0^-/1^-$ 
mesons described by Eqs.~(\ref{pseudf}) and (\ref{pseudg}) (or equivalently 
(\ref{fvec}) and (\ref{gvec})), and the orbitally excited $1^-$ meson 
described by Eqs.~(\ref{hvec}) and (\ref{kvec}).  Also indicated in Fig.~3 
are values of $\hat{\delta}_{\rm max}$ for each value of $\hat{\mu}$.  
As $\hat{\delta}$ approaches $\hat{\delta}_{\rm max}$ it is clear that 
the numerical calculation breaks down as the integration routine is unable 
to cope with the singularity in the integrand.  It is also clear that 
any reasonable extrapolation of the smooth part of the curves would not 
give a solution to $\Lambda(\delta) = 1$ for any of the curves in the range 
$\delta < \delta_{\rm max}$.  It also appears that carrying through the 
calculations for higher values of $\hat{\mu}$ would be unlikely to improve 
the situation.  We therefore conclude that, if the model is taken in its 
simplest form, namely the combination of bare quark-gluon vertex and 
Gaussian gluon propagator in a Feynman-like gauge, spurious poles encountered 
in solutions to the DSE prevent the existence of solutions to the meson BSE.  
If the model is to be applied to the physics of $D$ and $B$ mesons, less 
crude approximations must be employed.   

\setcounter{equation}{0}
\section{Conclusions.}

We have developed a non-perturbative treatment of heavy quarks which 
leads in a natural way to a dynamically generated self energy contribution 
to the heavy quark propagator.  
The treatment is based on a description of hadronic phenomena 
in terms of approximate Dyson-Schwinger equations which has proved highly 
successful within the light meson sector, namely the DSE technique 
\cite{RW94}).   

The extension of the DSE technique to heavy quarks is non-trivial.  
The heavy quark self energy must be calculated near the bare current quark 
mass pole $p^2 = -m_Q^2$, where $p$ is the Euclidean momentum.  This 
is accomplished by a change of momentum integration variable in the 
Dyson-Schwinger equation and an expansion in $1/m_Q$.  The dressed 
heavy quark propagator is then used in conjunction with light quark 
propagators obtained from the conventional DSE technique to obtain 
a ladder approximation Bethe-Salpeter equation for $D$ and $B$ mesons 
to leading order in $1/m_Q$.  The BSE gives the meson masses in the form 
$m_Q + \delta$, where $m_Q$ is essentially infinite.  At this level of 
approximation, the predictive power of the model lies not in calculating 
absolute meson masses, but in calculating mass differences between different 
spin states or different light quark content.  We recover the degeneracy 
between pseudoscalar and vector mesons predicted by heavy quark effective 
theory \cite{N94} in the limit $m_Q \rightarrow \infty$.  

We have carried out a numerical analysis of the model in the limit of 
infinite mass heavy quark, $m_Q \rightarrow \infty$, and massless 
light quark, $m_q = 0$.  Our calculation is not intended as a serious 
attempt to fit the $D$ and $B$ meson spectrum.  Instead it is a 
feasibility study to determine to what extent the method is computationally 
viable.  Solution of the BSE is numerically demanding, and the choice 
of model gluon propagator employed herein is primarily designed to 
reduce the numerical integration from three to two dimensions.  
We therefore use as input a particularly simple model gluon propagator, 
namely a gaussian propagator in a Feynman-like gauge.  Because we are 
working in the chiral limit of the light quark, all dimensionful parameters 
can be scaled by the the width of the gaussian function, and the model 
is defined  by a single adjustable parameter.  

We find the analytic structure of the heavy quark meson is consistent with 
that required for the change of integration variable used in the DSE.  
However we cannot unambiguously say whether the DSE leads to a confined 
heavy quark.  Our bare vertex approximation to the DSE yields a succession 
of masslike poles which move off the timelike axis one by one as the 
parameter in the model gluon propagator is varied.  As we point out below, 
we believe that these poles are a spurious artifact of the approximations 
employed.  

The combination of bare vertex and gaussian gluon propagator leads to 
poles in both the light and heavy quark propagators in the 
timelike half of the complex momentum plane over the broad range of input 
parameters studied.  While poles off the real timelike momentum axis do not 
in principle spoil quark confinement, they can prevent successful solution 
of the BSE, which samples the quark propagators throughout a region of the 
complex plane.  For the simple model set out above this is indeed the 
case.  We have demonstrated numerically that the poles 
generated by the bare vertex approximation restrict the meson mass 
$m_Q + \delta$ to values of $\delta$ which do not include any bound state 
solutions.  

In order to proceed further it will be necessary to develop an improved 
model which is not plagued by spurious singularities.  In ref.~\cite{BRW92} 
it is demonstrated that if the bare quark gluon vertex is replaced by 
a vertex ansatz satisfying the Ward-Takashi identity, such as the 
Ball-Chiu vertex \cite{BC80}, the quark propagator becomes an entire 
function.  This result holds independently of the current quark mass, 
and so must also hold in the heavy quark limit.  

Ideally one would like to incorporate quark propagator which is an entire 
function into our heavy quark formalism in a way consistent with 
successful DSE technique studies of light quark mesons.  
Of particular importance to the light meson sector is the requirement 
that the model should produce a Goldstone pion in the chiral 
limit.  This is achieved by ensuring that the choice of Bethe-Salpeter 
kernel is compatible with the choice of quark-gluon vertex in the DSE 
\cite{B96}.  While there is no practical way at present to achieve such 
a compatibility with the Ball-Chiu vertex, good results can be achieved 
for the light pseudoscalar and vector meson octets by inverting a bare 
vertex approximation DSE from an entire light quark propagator to construct a 
separable approximation to the Bethe-Salpeter kernel \cite{BSp1,C95}.  
Extension of the separable approximation in a similar manner 
to the heavy quark sector, 
beginning with the light quark propagators employed in ref.~\cite{BSp1} 
and a heavy quark propagator obtained from the Ball-Chiu vertex is 
perhaps the most promising way to proceed at this point.  One potentially 
then has a single non-perturbative model of meson dynamics applicable to 
the entire range of meson masses.  

Finally we note that, if the DSE technique is taken as a serious description 
of hadronic phenomena, we are led unavoidably to a treatment generically 
of the kind set out in this paper.  The above analysis illustrates for 
the first time the importance of the analytic structure of heavy quark 
propagators and its significance in determining the heavy meson spectrum.  
Our analysis leads to a heavy quark self energy 
which will contribute as significantly to the dynamics of the $D$ and $B$ 
mesons as that of the light quarks, and therefore cannot 
be ignored.  Furthermore we are unaware of any existing treatment of the heavy 
quark self energy which is genuinely non-perturbative and has the potential 
to provide a qualitative explanation of heavy quark confinement.  
For these reasons we believe it is important to persevere with the 
application of the DSE technique to the heavy quark sector.  

\section*{Acknowledgements.} 

The authors are grateful to the National Centre for Theoretical Physics 
at the Australian National University and the Institute for Theoretical 
Physics at the University of Adelaide for hospitality during visits where 
part of this work was completed.  We are happy to acknowledge C.-S.\ Huang 
for bringing the work of himself and his colleagues to our attention.  
One of us (D.-S.\ L.) would like to thank the Australian Research Council 
for their financial support under grant number A69231484. 

\renewcommand{\theequation}{\Alph{section}.\arabic{equation}}
\setcounter{equation}{0}
\setcounter{section}{1}
\section*{Appendix.}

We derive here a set of coupled integral equations from the DSE for the 
heavy quark propagator to order $1/m_Q$.  From Eqs.~(\ref{genprop}), 
(\ref{SABDEF}) and (\ref{kdef}) we obtain 
\begin{eqnarray}
\lefteqn{S(p) =
\frac{1 + \! \not \! v}{2}\frac{1}{ik\cdot v + \Sigma_B - \Sigma_A} +}
                             \label{SEXP} \\ 
& &  \!\!\!\!\!\!
   \frac{1}{2m_Q} \left[ \frac{\Sigma_B + \not \! v\Sigma_A - i  \not \! k}
       {ik\cdot v + \Sigma_B - \Sigma_A} -
    \frac{1 + \!\not \! v}{2}
            \frac{4ik\cdot v \Sigma_A + \Sigma_B^2 - \Sigma_A^2 + k^2}
       {(i k\cdot v + \Sigma_B - \Sigma_A)^2} \right] 
         + O\left(m_Q^{-2}\right),\nonumber
\end{eqnarray}
where, on choosing $v_\mu = (1,{\bf 0})$,
\begin{equation}
\Sigma_A = \Sigma_A(k_4,\left| {\bf k} \right|),
\hspace{5 mm}\Sigma_B = \Sigma_B(k_4,\left| {\bf k} \right|).
\end{equation}
Clearly $\Sigma_A$ and $\Sigma_B$ should depend only on one variable, 
namely
\[
p^2 = -m_Q^2 + 2im_Q k_4 + {k_4}^2 + \left| {\bf k} \right|^2.
\]
It follows that  
\begin{eqnarray}
\Sigma_A(k_4,\left| {\bf k} \right|) & = & \Sigma_A
   \left(k_4 - \frac{i}{2m_Q}\left| {\bf k} \right|^2 + 
                   O\left(m_Q^{-2}\right),0 \right)  \nonumber \\
 & = & \Sigma_A(k_4,0) - \frac{i}{2m_Q}\left| {\bf k} \right|^2
      \frac{\partial \Sigma_A(k_4,0)}{\partial{k_4}}
             + O\left(m_Q^{-2}\right). 
\end{eqnarray}

We now make the expansion
\begin{equation}
\Sigma_A(k_4,0) = \Sigma_A^0(k_4) + \frac{1}{m_Q} \Sigma_A^1(k_4) + 
       \ldots .
\end{equation}
Then 
\begin{eqnarray}
\Sigma_A(k_4,\left| {\bf k} \right|) & = & \Sigma_A^0(k_4) + 
 \frac{1}{m_Q} \left(\Sigma_A^1(k_4) - 
         \frac{i}{2}\left| {\bf k} \right|^2 {\Sigma_A^0}'\right)
             + O\left(m_Q^{-2}\right)  \nonumber \\
& = & \Sigma_A^0(k_4) + 
           \frac{1}{m_Q} \tilde{\Sigma}_A^1(k_4,\left| {\bf k} \right|)  
             + O\left(m_Q^{-2}\right), 
\end{eqnarray}
and similarly for $\Sigma_B$.  ($\tilde{\Sigma}_A^1$ is only defined for 
convenience here.  Eventually we shall be solving for $\Sigma_A^0$, 
$\Sigma_B^0$, $\Sigma_A^1$ and $\Sigma_B^1$, which only depend on $k_4$.)  
For the propagator Eq.~(\ref{SEXP}) we then have 
\begin{eqnarray} 
\lefteqn{S(p) =
 \frac{1 + \! \not \! v}{2}\frac{1}{ik_4 + \Sigma_B^0 - \Sigma_A^0} +
   \frac{1}{2m_Q} \left[ \frac{\Sigma_B^0 + \not \! v\Sigma_A^0 - i  \not \! k}
       {ik_4 + \Sigma_B^0 - \Sigma_A^0} - \right. } \nonumber \\
& &  \left.   \frac{1 + \!\not \! v}{2}
   \frac{4ik_4 \Sigma_A^0 + (\Sigma_B^0)^2 - 
                     (\Sigma_A^0)^2 + {k_4}^2 + \left| {\bf k} \right|^2
               +2\left( \tilde{\Sigma}_B^1 - \tilde{\Sigma}_A^1\right)}
       {(i k_4 + \Sigma_B^0 - \Sigma_A^0)^2} \right]
         + O\left(m_Q^{-2}\right).    \label{prpexp}
\end{eqnarray}

The left hand side of the DSE Eq.~(\ref{SDEqn}) then becomes 
\begin{eqnarray}
\lefteqn{\frac{i\!\not\! p}{m_Q} \Sigma_A + \Sigma_B = }\nonumber \\
 & & \!\!\! - \not\! v \Sigma_A^0(k_4) + \Sigma_B^0(k_4)
  +\frac{1}{m_Q} \left[i \!\not\! k  \Sigma_A^0(k_4) - 
     \!\not\! v \tilde{\Sigma}_A^1(k_4,\left| {\bf k} \right|) + 
                           \tilde{\Sigma}_B^1(k_4,\left| {\bf k} \right|)
             \right].
\end{eqnarray}
On the right hand side, we substitute in Eq.~(\ref{prpexp}) and change the 
variable of integration to $k'$ defined by $q_\mu = im_Q v_\mu + k'_\mu$.  
Then setting $k_\mu = (k_4,{\bf 0})$, and projecting out 
coefficients of $\not \!v$, 
$\not \! k$ and $I$ gives the following set of integral equations:
\begin{equation}
\Sigma_A^0(k_4) = \frac{4}{3}  \int \, \frac{d^4k'}{(2\pi)^4} \, 
    \Delta(k - k') \frac{1}{i k^\prime_4 + \Sigma_B^0(k^\prime_4)
          - \Sigma_A^0(k^\prime_4)} \, ,
                                \label{a7}
\end{equation}
\begin{equation}
\Sigma_B^0(k_4) = 2\Sigma_A^0(k_4), 
                                \label{a8}
\end{equation}
\begin{eqnarray}
\lefteqn{\Sigma_A^1(k_4) = \frac{4}{3}  \int \, \frac{d^4k'}{(2\pi)^4} 
   \, \Delta(k - k') \left[ 
    \frac{\Sigma_A^0}{i k^\prime_4 + \Sigma_A^0} - \right. }\nonumber \\
 & & \left. \left. \frac
   {4i k^\prime_4 \Sigma_A^0 + 3(\Sigma_A^0)^2 + 
                    (k^\prime_4)^2 + \left| {\bf k'} \right|^2 
 + 2\left(\Sigma_B^1 - \Sigma_A^1 - 
             \frac{i}{2} \left| {\bf k'} \right|^2 {\Sigma_A^0}'\right)}
     {(i k^\prime_4 + \Sigma_A^0)^2}\right]\right|_{k'},
\end{eqnarray}
\begin{eqnarray}
\lefteqn{\Sigma_B^1(k_4) = \frac{8}{3}  \int \, \frac{d^4k'}{(2\pi)^4} 
     \, \Delta(k - k') \left[ 
    \frac{2 \Sigma_A^0}{i k^\prime_4 + \Sigma_A^0} - \right.} \nonumber \\
 & & \left. \left. \frac
   {4i k^\prime_4 \Sigma_A^0 + 3(\Sigma_A^0)^2 + 
                                (k^\prime_4)^2 + \left| {\bf k'} \right|^2 
 + 2\left(\Sigma_B^1 - \Sigma_A^1 - 
              \frac{i}{2} \left| {\bf k'} \right|^2 {\Sigma_A^0}'\right)}
     {(i k^\prime_4 + \Sigma_A^0)^2}\right]\right|_{k'}.  
\end{eqnarray}
Once a model quark propagator $\Delta(k - k')$ is specified, these equations 
can, in principle, be solved numerically to give the heavy quark propagator 
to first order in $1/m_Q$.

%

%
\begin{table}
\caption{ \label{tabpole} Position of the poles $s^{\rm pole}$ in the 
light quark propagator and $k_4^{\rm pole}$ in the heavy quark 
propagator, and the maximum accessible values of $\delta_{\rm max}$ and 
corresponding momentum partitioning parameter $\eta_{\rm opt}$.  
Dimensionful quantities can be extracted from the dimensionless quantities 
listed via Eqs.~(\ref{scale1}) to (\ref{scale3}).} 

\begin{center}
\begin{tabular}{|ccccc|}\hline
       $\hat{\mu}$  &    $\hat{s}^{\rm pole}$ & $\hat{k}_4^{\rm pole}$ &
                  $\hat{\delta}_{\rm max}$ & $\eta_{\rm opt}$ \\ \hline
     2.5   &   $-0.126 + 0.000i$~  &  $1.802 + 0.957i$  &  1.307 & 0.268 \\
     3.0   &   $-0.540 + 0.000i$~  &  $2.285 + 1.034i$  &  1.769 & 0.415 \\
     3.5   &   $-1.829 + 0.919i$~  &  $2.769 + 1.102i$  &  2.494 & 0.558 \\
     4.0   &   $-1.646 + 2.234i$~  &  $3.255 + 1.162i$  &  2.649 & 0.561 \\
     5.0   &   $-1.246 + 4.546i$~  &  $4.229 + 1.264i$  &  2.991 & 0.577 \\
     6.0   &   $-0.327 + 7.216i$~  &  $5.207 + 1.351i$  &  3.294 & 0.589 \\
     7.5   &   ~$2.132 + 11.711i$  &  $4.680 + 2.649i$  &  4.859 & 0.455 \\ 
\hline
\end{tabular}
\end{center}
\end{table}
\pagebreak
\section*{Figure captions.}
\begin{description}
 \vspace*{5mm}
  \item[Figure 1:] The heavy quark self energy obtained by solving 
Eq.~(\ref{sigie}): ${\rm Re}\hat{\Sigma}(\hat{k}_4)$ (dashed curve) and  
$-{\rm Im}\hat{\Sigma}(\hat{k}_4)$ (solid curve) for 
$\hat{\mu} =  \mu/\sqrt{\alpha} = 3.5$.  
  \vspace*{5mm}
   \item[Figure 2:] The position $\hat{k}_4^{\rm pole}$ of the poles in 
the heavy quark propagator for the values of $\hat{\mu}$ indicated.  
Only poles for which ${\rm Re}\,k_4 > 0$ are shown, though the poles 
come in pairs $\hat{k}_4^{\rm pole}$ and $-\hat{k}_4^{\rm pole\,*}$
  \vspace*{5mm}
   \item[Figure 3:] Eigenvalues $\Lambda(\hat{\delta})$ from 
Eq.~(\ref{lineq}) for the degenerate $0^-/1^-$ states (solid curves) and 
the orbitally excited $1^-$ state (dashed curve) calculated with 
gluon propagator parameter $\hat{\mu} = 3.5$ (triangles), 5.5 (circles) 
and 6.0 (squares).  Also shown as vertical dotted lines are the numerical 
limits $\hat{\delta}_{\rm max}$ for each value of $\hat{\mu}$.  
\end{description}
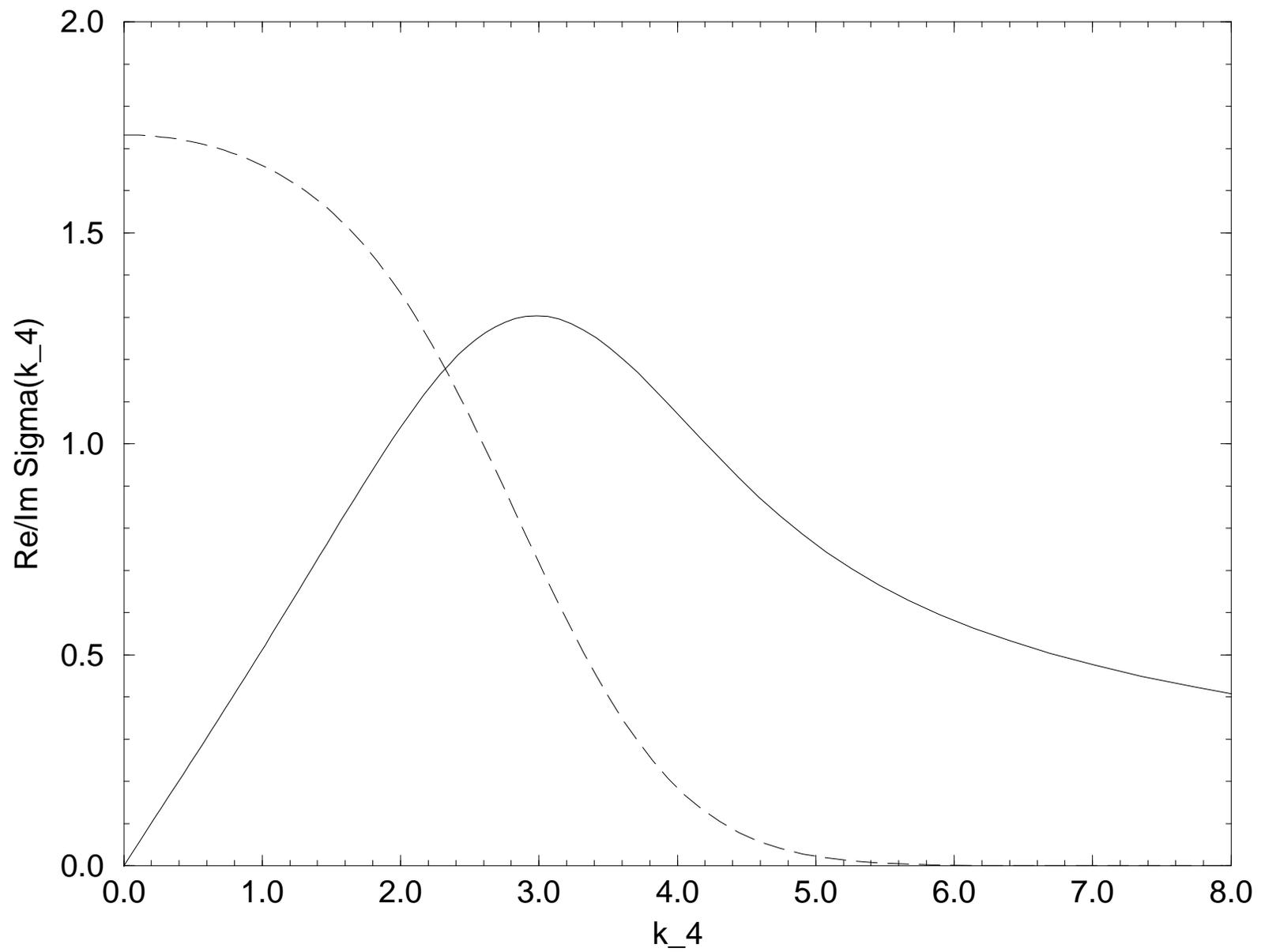
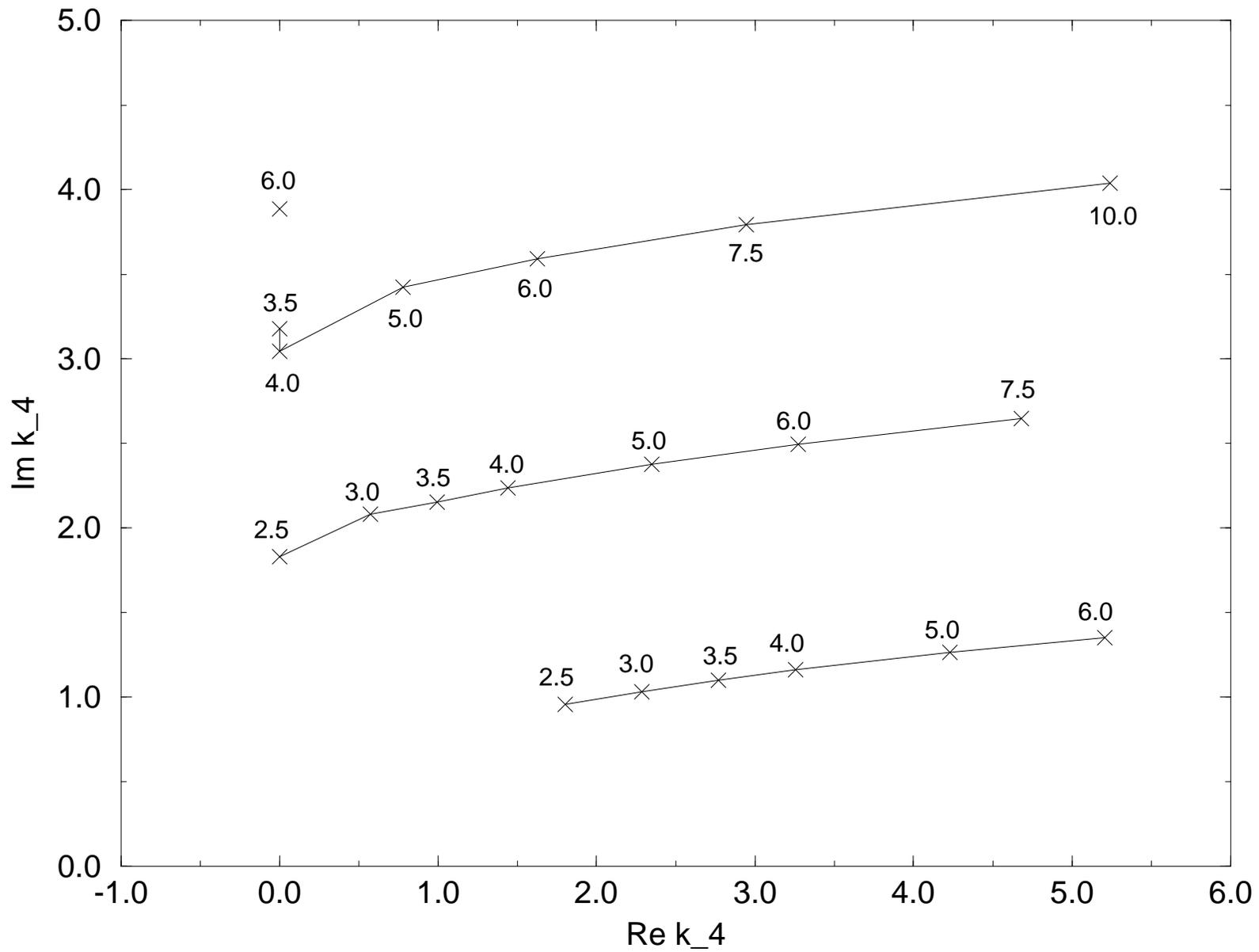
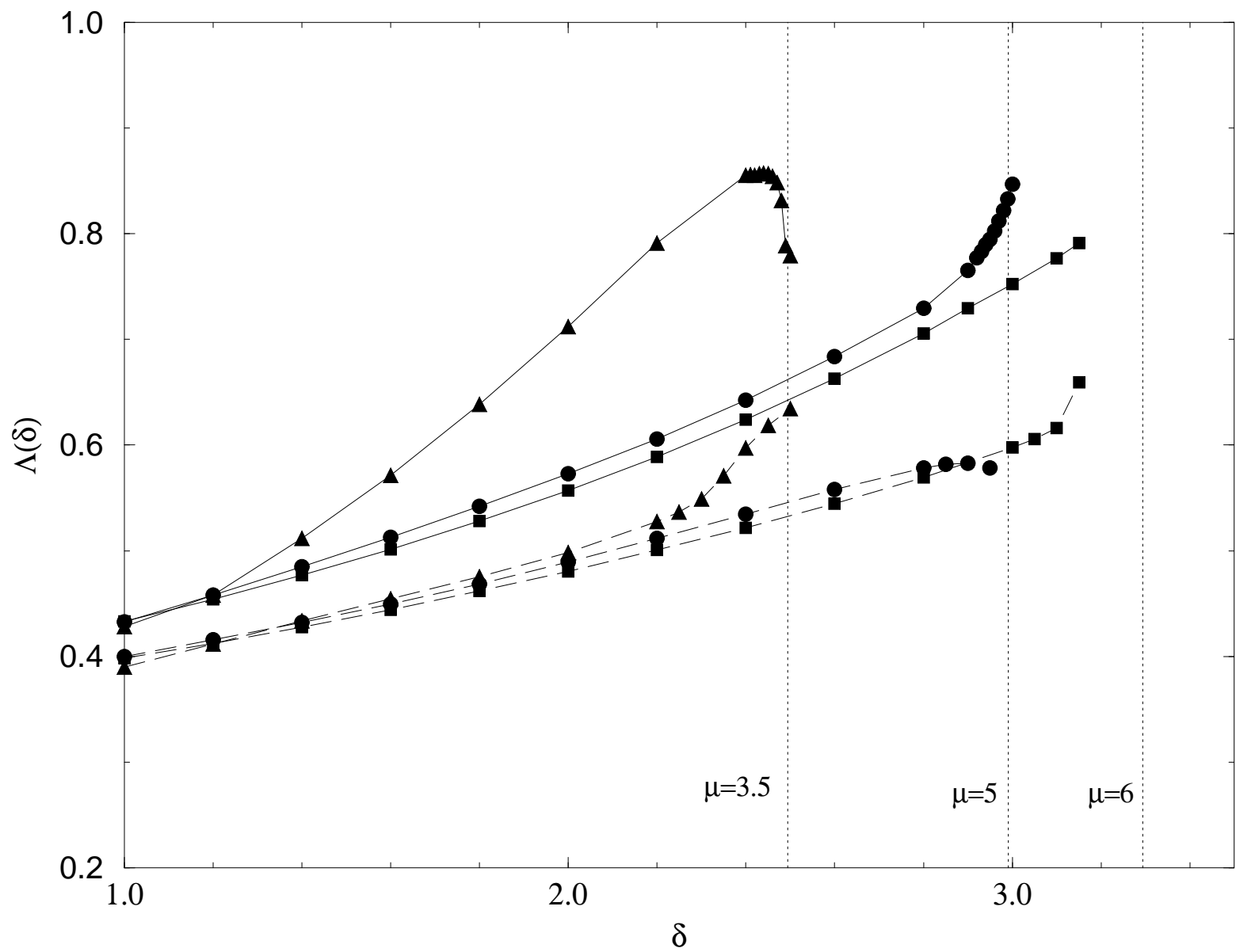
\end{document}